\documentstyle[epsf]{elsart}

\begin{document}
\begin{frontmatter}

\title{Gauge theory of self-similar system}

\author{Alexander I. Olemskoi\thanksref{AOVB}}
\address{Physical Electronics Department, Sumy State University\\
2, Rimskii-Korsakov St., 40007 Sumy UKRAINE}

\thanks[AOVB]{olemskoi\char'100ssu.sumy.ua, alexander\char'100olem.sumy.ua}

\begin{abstract}

On the basis of a dilatation invariant Lagrangian,
governed equations are determined for probability density
and gauge potential of the non-stationary self-similar stochastic system.
It is shown that an automodel regime is observed
at small time interval determined by the Tsallis' parameter $q>1$.
An exponential falling down happens at large time where
the dilatation parameter and the partial scale tend to constant values.

\vspace{0.5cm}


\vspace{0.5cm}


\end{abstract}

\end{frontmatter}

\section{Introduction}

Let us consider the one-dimensional random walker with coordinate $x$
at time $t$ -- a characteristic example of the stochastic system
carrying out L\'evy flights \cite{1}.
In the case of self-similar system, the corresponding probability
distribution $P(x, t)$ is a homogeneous function satisfying
to the condition
\begin{equation}
P(x, t)=a^{-\alpha}{\cal P}(\kappa),\quad
\kappa\equiv x/a
\label{1}
\end{equation}
where $a\equiv a(t)$ is a time dependent partial scale,
$\kappa$ is a dimensionless coordinate,
$\alpha$ is a self-similarity index.
To analyze such a system, it is convenient to use the basic conception
of Jackson's derivative ${\cal D}_q$, whose properties are given in Appendix.
Basic advantage of this  derivative for analyzing self-similar system
is that the Jackson's derivative determines
rate of the function variation with respect to dilatation $q$,
but not to the shift ${\rm d}x\to 0$ as in usual case.
This work is devoted to studying a stochastic self-similar system
on the basis of such a type representation.

The paper is organized as follows. In Section 2 the governed equations
for probability density and gauge potential are obtained
starting from a dilatation invariant Lagrangian.
Section 3 deals with the determination of
the time dependencies for a characteristic scale and probability density.
A steady state is shown to realize when the Jackson's derivative
of the gauge potential equals to zero. 
This represents a gauge condition,
under which the probability distribution has the Tsallis' form.
With breaking the gauge condition, an automodel regime is observed
at small time interval determined by the Tsallis' parameter $q>1$.
An exponential falling down happens at large time where
the dilatation parameter and the partial scale tend to a constant values.
Section 4 contains a short conclusions and in Section 5 
basic properties of the Jackson's derivative are adduced.

\section{Basic equations}

In contrast to the simple case, when the scale $a$
does not depend on the time $t$, we study here
a non-stationary self-similar system, for which the value $a$
and the dilatation factor $q$ are functions of $t$.
As it is known from the theory of the gauged fields \cite{2},
in such a case the system is invariant with respect to transformations
$x\to q x$, $P\to Q_q P$, $Q_q\sim q^{\alpha}$
if the gradient terms ${\cal D}_q P$, $\partial P/\partial t$
are replaced by the elongated derivatives $({\cal D}_q+\epsilon)P$,
$(\partial/\partial t + E)P$
with dilatational $\epsilon$ and temporal $E$ components
of the gauge potential (hereafter the time $t$ is measured
in units of the probability relaxation time).
In accordance with Eq.(\ref{4a}), it is easy to see
that this elongated derivatives are invariant
with respect to the non-stationary dilatation $q=q(t)$
determined by the follow transformations:
\begin{eqnarray}
&x\to q x,\qquad P\to Q_qP;&\nonumber\\
\label{2}\\
&\epsilon\to\epsilon-{{\cal D}_q Q_q\over Q_q}-
(q-1){{\cal D}_q Q_q\over Q_q}~{{\cal D}_q P\over P},\qquad
E\to E-\dot Q&
\nonumber
\end{eqnarray}
where point denotes the time derivative for brevity.

Gauge invariant Lagrangian of the corresponding Euclidean field theory
is supposed to take the form
\begin{equation}
{\cal L}={1\over2}\left[({\cal D}_q+\epsilon)P\right]^2+
{1\over2} ({\cal D}_q \epsilon)^2
\label{3}
\end{equation}
where the first term is caused by the gauged dilatation, the second one is
the field contribution.
The respective dissipative function reads
\begin{equation}
f={1\over2}\left[\left({\partial\over\partial t}+E\right)P\right]^2+
{\theta\over2}\left[\left({\partial\over\partial t}+E\right)\epsilon\right]^2
\label{4}
\end{equation}
where $\theta$ is the relaxation times ratio of the gauge field
and probability.
As a result, the Euler equation
\begin{equation}
{\cal D}_q {\partial{\cal L}\over\partial ({\cal D}_q {\rm z})}-
{\partial{\cal L}\over\partial{\rm z}}=
-{\partial f\over\partial\dot{\rm z}},\quad {\rm z}\equiv (P,\epsilon)
\label{5}
\end{equation}
leads to the differential equations with partial derivatives
and non-linear terms:
\begin{eqnarray}
\dot P + {\cal D}_q^2 P &=& -EP+\epsilon^2 P,\label{6}\\
\theta\dot\epsilon+{\cal D}_q^2 \epsilon-P {\cal D}_q P &=&
-\theta EP+\epsilon P^2.
\label{7}
\end{eqnarray}
The first terms in right-hand parts describe
dissipative influence of external environment.
Further, we will take into consideration conserved systems only,
so that the time component of the gauge potential
being inversely proportional to corresponding relaxation time
will be put equal to zero $(E=0)$.

\section{Solution of equations}

In the limits $\epsilon\to 0$, ${\cal D}_q^2 \epsilon\to 0$, 
the obtained equations (\ref{6}), (\ref{7}) take the form
\begin{eqnarray}
\dot P &=& -{\cal D}_q^2 P,\label{8}\\
\theta\dot\epsilon&=&P {\cal D}_q P.
\label{9}
\end{eqnarray}
The first of them has the diffusion type but with inverted sign,
so that self-similar system reveals running away kinetics
that is inherent in hierarchical systems \cite{4}.
However, such a behaviour realizes during short time interval
$t\sim\theta\ll 1$ only.
At usual time $t\ge 1$, we can use the condition $\theta\ll 1$
of adiabatic approximation that
will be used everywhere below and allow to
neglect left-hand side of Eq.(\ref{9}).
As a result, the condition ${\cal D}_q P\approx 0$ holds true and
the system passes to stationary homogeneous regime:
\begin{equation}
P(x,t)=const\equiv P_{st} .
\label{10}
\end{equation}
In much more complicated limit $\epsilon(t)\to const\ne 0$,
equation (\ref{7}) is reduced to the simplest form
\begin{equation}
{\cal D}_q P =-\epsilon P
\label{11}
\end{equation}
and we arrive at the static distribution (\ref{10}) as before.

To continue analysis, let us multiply
Eq.(\ref{6}) by factor ${\cal D}_q P$ and Eq.(\ref{7}) by ${\cal D}_q \epsilon$.
Then, after addition of the obtained results we find
\begin{equation}
{1\over2}({\cal D}_q P)^2+{1\over2}({\cal D}_q\epsilon)^2 =
{1\over2}(\epsilon P)^2+{1\over2}C^2.
\label{12}
\end{equation}
Here we put
\begin{equation}
(\dot P - P{\cal D}_q\epsilon){\cal D}_q P=0
\label{13}
\end{equation}
and fulfilled integration with constant $C^2/2$.
As a result, under the assumption
\begin{equation}
{\cal D}_q\epsilon=-C,\quad C>0
\label{14}
\end{equation}
and condition ${\cal D}_q P\ne 0$,
the equation (\ref{13}) arrives at the exponential time dependence
\begin{equation}
P\propto {\rm e}^{-Ct}
\label{15}
\end{equation}
whereas the equation (\ref{12}) is reduced to the form (\ref{11}).

The exponential falling-down is known to
be not inherent in the self-similar systems \cite{3} and,
as a consequence, we are need to put $C=0$ in Eqs.(\ref{12}), (\ref{14}).
We arrive then at the gauge condition
\begin{equation}
{\cal D}_q\epsilon=0,
\label{16}
\end{equation}
according to which the potential $\epsilon$ can be
a time dependent function
but does not vary with the system dilatation.
Then, equation (\ref{6}) takes the form $\dot P=0$
meaning that the system is in steady-state, which
probability distribution is obeyed to Eq.(\ref{11}).
Being accompanied with  Eq.(\ref{2a}), this equation arrives at
the condition $[\alpha]_q=-\epsilon$.
As is ascertained in Appendix,
the Jackson's q-number $[\alpha]_q$ is reduced
to the Tsallis' q-logarithm if the steady-state probability $P_{st}$
and dilatation $q$ are connected via the follow relation:
\begin{equation}
P_{st}^{q-1}\equiv q^{\alpha}.
\label{17}
\end{equation}
Then, the above obtained condition
gives the Tsallis' distribution \cite{3}
\begin{equation}
P_{st}=[1-(q-1)\epsilon]^{1\over q-1}.
\label{18}
\end{equation}

With breaking gauge condition (\ref{16}) the self-similar system
gets into non-stationary state, whose behaviour
is determined by Eq.(\ref{6}).
Accounting definitions (\ref{5a}) arrives this to the algebraic form
with respect to the Jackson's derivation:
\begin{equation}
\dot P =  \left(\epsilon^2-[\alpha]_{qq}\right) P.
\label{19}
\end{equation}
Inserting here Eq.(\ref{1}) and taking into
consideration the condition $[\alpha]_q=-\epsilon$ and relation
$\dot P = -a^{-(1+\alpha)}(\alpha{\cal P}+\kappa{\cal P}')\dot a$
(hereafter prime denotes the usual derivation with respect
to the argument $\kappa$) we obtain
\begin{equation}
a^{-1}\dot a(\alpha{\cal P}+\kappa{\cal P}')=[\delta\alpha]_{qq}{\cal P} 
\label{20}
\end{equation}
where the factor $[\delta\alpha]_{qq}$ stands for 
the term determined by Eqs.(\ref{5a}). 

In the limit $q\to\infty$, one can see with accounting asymptotics (\ref{6a})
that system behaves in automodel manner
if conditions $aq=const$, $a^{3\alpha-4}\dot a = const\equiv\tau_0^{-1}$
and equation
\begin{equation}
\kappa{\cal P}'=(\tau_0-\alpha){\cal P}
\label{21}
\end{equation}
are implemented.
Solution of the equation is ${\cal P}\propto\kappa^{\tau_0-\alpha}$
and the time dependencies of the characteristic scale and
the probability density read:
\begin{equation}
a^{3(\alpha-1)}={t\over\tau}, \quad
P\propto x^{\tau_0-\alpha}~t^{-\tau},~
\tau\equiv{\tau_0\over 3(\alpha-1)}
\quad {\rm at} \quad q\gg 1,~~t<\tau.
\label{22}
\end{equation}

Within the opposite limit $q\to 1$, a magnitude $q$ in Eq.(\ref{20})
ought to put time independent and we arrive at
the long-time dependencies:
\begin{equation}
a\propto \exp{(t/\tau_0)}, \quad
P\propto x^{\lambda_0\tau_0-\alpha}\exp{(-\lambda_0 t)},~
\lambda_0\equiv{\alpha-1\over q-1}
\quad {\rm at} \quad q\to 1,~t\gg\lambda_0^{-1}.
\label{23}
\end{equation}
The coincidence condition for the time limits $\tau$ and $\lambda_0^{-1}$
in dependencies given by Eqs.(\ref{22}), (\ref{23})
leads to relation
\begin{equation}
\tau=3(q-1).
\label{24}
\end{equation}

At last, we consider the case with non-zeroth second Jackson's derivative
${\cal D}^2_q\epsilon = [\varepsilon]_{qq}\epsilon$
determined by an index $\varepsilon$.
Here equation (\ref{7}) gives the gauge potential
\begin{equation}
\epsilon = {-[\alpha]_q\over 1-[\varepsilon]_{qq}P^{-2}}
\label{25}
\end{equation}
that behaves in self-similar manner if the value
$[\varepsilon]_{qq}P^{-2}$ falls down with $q$-increase.
Supposing this falling down in power form $q^{-\gamma}$
with positive index $\gamma\to 0$, we obtain the needed
dependence $P(t)\propto q^{\tau}(t)$, following from Eqs.(\ref{22})
and condition $a(t)q(t)=const$, if $\gamma=2\tau-3(\varepsilon-1)>0$.
As a result, the gauge potential index $\varepsilon$
is limited by the condition
\begin{equation}
\varepsilon < 1 + {2\over3}~\tau = 1+2(q-1)
\label{26}
\end{equation}
where the second equality follows from Eq.(\ref{24}).
Under this conditions the equation (\ref{6}) accompanied with
approximated result $\epsilon\approx -[\alpha]_q$,
following from Eq.(\ref{25}),
arrives at the above obtained time dependencies (\ref{22}).

The automodel regime is studied to be broken if the factor
in right-hand side of Eq.(\ref{19})
\begin{equation}
\epsilon^2-[\alpha]_{qq}=
[\alpha]_q^{2}\left[(1-[\varepsilon]_{qq}P_{st}^{-2})^{-2}
-1\right]-[\delta\alpha]_{qq}\equiv-\lambda
\label{27}
\end{equation}
becomes time independent (here
the steady-state probability $P_{st}$ is determined by Eq.(\ref{18}) and
we take into account Eqs.(\ref{25}), (\ref{5a})).
In such a case, the exponential decay (\ref{23}) is
characterized by the relaxation time $\lambda^{-1}$
instead of $\lambda_0^{-1}$.
Because this regime is inherent in small values of $q$, we can use
the limit $q\to 1$:
\begin{equation}
\lambda=\lambda_0-\alpha^{2}
\left\{\left[1-\left(\varepsilon^2+{\varepsilon-1\over q-1}\right)
{\rm e}^{2\epsilon}\right]^{-2}-1\right\},\quad
\lambda_0\equiv{\alpha-1\over q-1}.
\label{28}
\end{equation}
Finally, under conditions $q\to 1$,
$\varepsilon^2+(\varepsilon-1)/(q-1)\ll{\rm e}^{2\alpha}$
when $\alpha=[\alpha]_q=-\epsilon$, we have
\begin{equation}
\lambda=\lambda_0\left[1-2\alpha^2{q-1\over\alpha-1}
\left(\varepsilon^2+{\varepsilon-1\over q-1}\right)
{\rm e}^{-2\alpha}\right].
\label{29}
\end{equation}

\section{Discussion}

The above offered formalism is based on the dilatation invariant
Lagrangian (\ref{3}) and dissipative function (\ref{4}) to describe
the conserved non-stationary self-similar stochastic system.
Behaviour of such a system is determined by the probability density
distribution (\ref{1}) and the gauge potential $\epsilon$ (the latter 
is reduced to the ratio of the microstate energy to temperature 
in usual case of thermodynamic systems).
The non-linear differential equations with partial derivatives,
Eqs.(\ref{6}), (\ref{7}) are obtained to analyze the system kinetics.
It is occurred that under gauge condition (\ref{16}), when 
the microstate energy is independent on dilatation, the system is in
the steady-state characterized by the Tsallis' distribution (\ref{18}).
With gauge breaking, the automodel regime (\ref{22}) realizes 
for a time less than bounded magnitude (\ref{24}) determined
by the difference $q-1$. For more values of time, when the dilatation
becomes constant, the system passes to the usual exponential
regime (\ref{23}).

Finally, we comment on limitations of our approach. 
The main of these is that the system under consideration is conserved,
so that influence of external environment
have been put equal to zero (the value $E=0$ in Eqs.(\ref{6}), (\ref{7})).
Accounting such type terms for hierarchical systems shows that
the time component of the gauge potential is reduced
to the linear differential operator
$E\equiv -(\partial/\partial x)F(x) + (\partial^2/\partial^2 x)D(x)$
to be anti-dissipative in the physical meaning
(here $F(x)$ is a drift force and $D(x)$ is a diffusion coefficient) [5, 6].
Due to these terms, the above found exponential regime
will be suppressed and the self-similar system will behave in automodel manner
during whole time interval.

\section*{Acknowledgment}

I am grateful to Constantino Tsallis for sending his
review [3], which studying inspired me to this work.

\section{Appendix. Basic properties of the Jackson's derivative}

The Jackson's derivative is defined by equation
\begin{equation}
{\cal D}_q f(x)\equiv {f(qx)-f(x) \over q-1},\quad q\ne 1
\label{1a}
\end{equation}
that is reduced to the usual derivative in the limit $q\to 1$.
Apparently, for a homogeneous function
with a self-similarity index $\alpha$
the Jackson's derivative is reduced to Jackson's q-number $[\alpha]_q$:
\begin{equation}
{\cal D}_q f(x)=[\alpha]_q f(x),\qquad
[\alpha]_q\equiv {q^{\alpha}-1 \over q-1}.
\label{2a}
\end{equation}
It is easily to see that the value $[\alpha]_q\to\alpha$ in the limit
$q\to 1$ and increases as $q^{\alpha-1}$
at $q\to \infty$ (we propose $\alpha>1$).
On the other hand, the Tsallis' q-logarithm function
$\ln_q x\equiv(x^{q-1}-1)/(q-1)$ can be represented in the form
of the  Jackson's q-number with index
$\alpha=(q-1)\ln x/\ln q$.
Accompanied Eq.(\ref{2a}) this relation and apparent equality \cite{3}
\begin{equation}
\ln_q(xy)=\ln_q x + \ln_q y + (q-1)(\ln_q x)(\ln_q y)
\label{3a}
\end{equation}
lead to important rule for the Jackson's derivative:
\begin{equation}
{\cal D}_q \left[f(x)g(x)\right]=\left[{\cal D}_q f(x)\right]g(x)+f(x)\left[{\cal D}_q g(x)\right]
+ (q-1)\left[{\cal D}_q f(x)\right]\left[{\cal D}_q g(x)\right].
\label{4a}
\end{equation}
The Jackson's derivative of the second order is determined as follows:
\begin{eqnarray}
&{\cal D}_p {\cal D}_q f(x)={\cal D}_p\{[\alpha]_q f(x)\}=[\alpha]_{pq}f(x),&
\nonumber\\
\label{5a}\\
&[\alpha]_{pq}\equiv [\alpha]_p [\alpha]_q +[\delta\alpha]_{pq},\quad
[\alpha]_{p}\equiv{p^{\alpha}-1\over p-1},\quad
[\delta\alpha]_{pq}\equiv{p^{\alpha}[(pq)^{\alpha}-pq]\over (p-1)(pq - 1)}.&
\nonumber
\end{eqnarray}
In proposition $\alpha>1$, the value $[\delta\alpha]_{pq}$
has the follow asymptotics:
\begin{eqnarray}
&[\delta\alpha]_{pq}\to {\alpha-1\over p-1}
\quad {\rm at} \quad p,q\to 1,&
\nonumber\\
\label{6a}\\
&[\delta\alpha]_{pq}\to p^{2(\alpha-1)}q^{(\alpha-1)}
\quad {\rm at} \quad p,q\to \infty.&
\nonumber
\end{eqnarray}


\begin{thebibliography}{00}

\bibitem{1} J.-P.~Bouchaud, A.~Georges, Phys. Rep.
{\bf 195}, 127 (1990).

\bibitem{2} J. Zinn-Justin, {\it Quantum Field Theory and
Critical Phenomena} (Clarendon Press, Oxford, 1993).

\bibitem{3} C. Tsallis, {\it Nonextensive statistical mechanics
and thermodynamics: Historical background and present status},
in {\it Nonextensive Statistical Mechanics and its Applications},
{\it  Lecture Notes in Physics}, eds. S. Abe and Y. Okamoto
(Springer-Verlag, Berlin, 2000).

\bibitem{4} A.~I.~Olemskoi, A.~D.~Kiselev,
Phys. Lett. A {\bf 247}, 221 (1998).

\bibitem{5} A.~I.~Olemskoi, JETP Letters {\bf 69}, 423 (1999).

\bibitem{6} A.~I.~Olemskoi, JETP Letters {\bf 71}, No.7 (2000).

\end{thebibliography}
\end{document}